\documentstyle[11pt]{article}
\begin{document}
\centerline{\Large\bf Relativistic restrictions on the distinguishability}
\vskip 2mm
\centerline{\Large\bf of orthogonal quantum states}
\vskip 2mm

\centerline{S.N.Molotkov and S.S.Nazin}
\centerline{\sl\small Institute of Solid State Physics 
of Russian Academy of Sciences}
\centerline{\sl\small Chernogolovka, Moscow District, 142432, Russia}

\begin{abstract}
We analyze the restrictions on distinguishability of quantum states
imposed by special relativity. An explicit expression relating the
error probability for distinguishing between two orthogonal single-photon 
states to the time interval $T$ between the beginning of the measurement 
procedure and the moment when the measurement result is obtained by the 
observer.
\end{abstract}
\vskip 2mm

Many problems of non-relativistic quantum information theory involve the task 
of distinguishing between two quantum states.

Basic principles of non-relativistic classical physics implicitly assume
that any measurement of a physical system can in principle be carried out with 
arbitrarily high accuracy and without disturbing the system state. Moreover, 
because of the absence of any restrictions on the maximum possible speed any 
measurement (even spatially non-local) can in principle be carried out 
arbitrarily fast (formally, in zero time). Therefore any two states of 
a physical system can be can be distinguished reliably, instantaneously, 
and without disturbing them.

In the non-relativistic quantum mechanics any measurement of a quantum system
generally disturbs its state. There is a fundamental difference between 
the distinguishing of a pair of orthogonal and a pair of non-orthogonal quantum
states. For orthogonal states the state of a quantum system can be reliably 
(with zero error probability) identified without disturbing it [1,2].
The very possibility of obtaining
the measurement result instantaneously (in zero time) implicitly contains 
the absence of any restrictions on the maximum possible speed.

Non-orthogonal states are fundamentally non-distinguishable reliably, i.e.,
it is never possible to distinguish between the two non-orthogonal 
states of a quantum system with zero error probability. The exact lower 
boundary for the probability of this error has been established long ago
[3--5]. That is why all the non-relativistic quantum cryptographic protocols
are based on non-orthogonal states. There exist no fundamental restrictions
on instantaneous distinguishing (although with non-zero error probability) 
between a pair of non-orthogonal states.

The relativistic quantum theory (which actually arises as the quantum field 
theory, since no meaningful interpretation of relativistic quantum mechanics
can be proposed) should also contain additional (compared with the 
non-relativistic quantum mechanics) restrictions on the time required for 
distinguishing of quantum states. The fundamental restrictions imposed by
special relativity on the measurability of dynamical 
observables of quantum systems were first considered in 1931 in the paper 
by Landau and Peierls [6]. Qualitative analysis of Ref. [6] based on the
uncertainty relations together with the limitation on the maximum possible
speed led to the conclusion that, in contrast to the non-relativistic quantum 
mechanics, in the relativistic theory the exact determination of momentum
can no longer be preformed in any finite time. The authors of Ref. [6]
actually arrive at the conclusion that no non-local dynamical variable of 
a quantum system can be measured.

In the non-relativistic theory the momentum can in principle be measured with
any accuracy in spite of the fact that the momentum eigenfunctions are the plane
waves which are infinitely extended in space. To be more precise, the plane 
wave is not a physically realizable state since it does not belong to the 
Hilbert space of square-integrable functions and instead can be considered as a
generalized eigenvector of the momentum operator [7] (i.e., a continuous linear
functional in the rigged Hilbert space [8]). Any generalized eigenvector 
(plane wave) can be approximated with arbitrarily high accuracy by a normalized
state localized in a large enough but still finite spatial domain yielding
the average value of the momentum operator equal to the plane wave momentum.
The momentum measurement assumes the accessibility of the entire state
extended over arbitrarily large spatial domain. In the non-relativistic
quantum mechanics there are no restrictions on having an instantaneous access
to any domain and hence the momentum (and other observables) can in principle
measured with arbitrarily high accuracy. However, special relativity implies
that the access to an infinite domain requires infinite time and in that sense 
the dynamical variables are cannot be measured if we demand their determination 
in a finite time.

The analysis of measurements of relativistic quantum systems was further
advanced in 1933 in the work of Bohr and Rosenfeld [9]. Their critical 
remarks concerning Ref.[6] do not affect the restrictions derived in that work
since they actually follow from the basic principles of special relativity.
and can only be eliminated rejecting the special relativity theory.
The arguments of Ref. [6] were later reproduced in Ref. [10] without any
changes. 

The orthogonality of two quantum states is, strictly speaking, a non-local 
property, both in Hilbert space and in the Minkowskii space-time. However,
this circumstance alone does not imply. for example, that the two orthogonal 
states cannot be distinguished by local measurements (in the sense that the
measurement outcome can be ascribed to a particular spatial point).

Special relativity imposes additional restrictions on the distinguishability
of the states of a physical system. These restrictions arise already in the
classical relativistic physics. Due to the existence of maximum possible 
speed, the instantaneous spatially non-local measurements become impossible. 
In the relativistic theory the very concepts of time in simultaneity are no
longer absolute and become meaningful only when related to a certain reference 
frame. However, the specific nature of the problems related to the transfer of
information implies that the events of sending (preparing a state) and 
receiving information (measuring the state) should be causally related, 
i.e. separated by a time-like interval which does not depend on the reference 
frame. 

The relativistic restrictions on non-local measurements arising already in 
the classical case can be seen from the following simple example. Suppose
the observer has to distinguish between two extended objects (two rulers
with different known lengths $L_1$ and $L_2$ aligned along 
the $x$-axis) which are randomly submitted to him. The length is defined in a 
standard way as the difference between the coordinates of the ruler ends taken
in the specified (laboratory) reference frame at the same moment of time [11].
Measurement of length implicitly assumes that the information should be 
transferred from the ruler ends to the observer which inevitable requires a 
finite time of at least $T=L_{min}/2c$ ($L_{min}=\min \{L_1,L_2\}$. 
More formally, the space-like cross-section of the extended objects (rulers)
should be fully covered by the backward light cone issued
from the point were the observer is located (Fig. 1).

Similarly, the time required for distinguishing between different 
three-dimensional classical objects is determined as the time
obtained when covering the largest cross-section (by a space-like
hyperplane) of the object at a given time by the back part of the light cone
(Fig. 1).

Change to a different inertial reference frame moving relative to the
laboratory one results in the Lorentz contraction of the geometrical sizes
by a factor of $\sqrt{1-\beta^2}$ and deceleration of time by the same factor.
However, the time measured with the clocks in the laboratory
reference frame remains unaltered, and it is this time that is important
for the information exchange protocols.

Hence, it is not surprising that distinguishing of infinitely extended
objects would require infinite time.

Bearing in mind problems of quantum information theory, we shall be 
interested in the restrictions imposed by the relativistic quantum theory
on the time required to obtain a measurement result when distinguishing
between two orthogonal states of a quantum system. Although the arguments
on the finiteness of the time required for the observer to obtain a final
measurement result are very general in nature, the time in question depends
on the particular structure of the states involved. 

{\it Below we shall consider
the problem of distinguishing between two orthogonal single-photon states
for which it is possible establish a general relation between the error of 
state identification and the time interval $T$ between the beginning of the 
measurement procedure and the moment when the measurement result is obtained 
by the observer.}

For our purposes, it is sufficient to consider only the pure states since
any state can be written (although generally not in unique way) as a 
statistical mixture of pure states.

Suppose we have a pair of orthogonal states in the Hilbert space
${\cal H}$, $|\psi_{0,1}\rangle\in{\cal H}$, $\langle\psi_0|\psi_1\rangle=0$.
There are three natural levels of description of the measurement process
in quantum mechanics which differ in the amount of information they provide
[4,12--15]. The simplest description of the measurement procedure lists
only the possible measurement outcomes (i.e., specifies the space of possible
measurement outcomes $\Theta$) and gives the relative frequencies 
(probabilities) of occurrence of particular outcomes (i.e. the probability 
of the measurement outcome lying in a measurable subset 
$\Delta \subset \Theta$) for a given input state of the
measured quantum system. In that sense the measurements are in one-to-one 
correspondence with the positive identity resolutions on $\Theta$ in $\cal H$ 
[3,4,12--15], i.e. the families of Hermitian operators 
$M_t(\Delta)$, $\Delta \subset \Theta$, which act in $\cal H$ and satisfy the 
following properties:\\
\noindent $1$) $M_t(\emptyset)=0$, $M_t(\Theta)=I$, (normalization)\\
\noindent $2$) $M_t(\Delta)\ge 0$, (positivity) and \\
\noindent $3$) $M_t(\Delta)=\sum_jM(\Delta_j)$, if
$\Delta=\cup_j \Delta_j$, $\Delta_j\cap \Delta_i=\emptyset$ for $i \neq j$
(additivity).\\
In this approach, the measure $\mu_{t,\rho}$ of a set $\Delta$ is defined as
\begin{equation}
\label{Prob_M} \mu_{t,\rho}(\Delta) = {\rm Prob}_t(\theta \in \Delta) =
\mbox{Tr}\{\rho M_t(\Delta)\}.
\end{equation}
In other words, $M_t(\Delta)$ define a positive operator-valued measure.
Different parameters $t$ correspond to physically different measurements.
Here the parameter $t$ has the meaning of the moment of time when the 
measurement began. It should be emphasized that the time $t$ is a parameter 
and does not belong to the outcome space. It has also nothing to do with the 
moment of time when the final measurement result is obtained by the observer 
(related to the time required to read off and deliver the classical information 
from a non-local measuring apparatus to the observer). To avoid confusion, we 
shall use the capital letter $T$ for the latter time. 

Specification of a positive operator-valued measure (POVM) is a formal 
description of a physical device realizing a black box taking a
quantum state as an input and producing a classical output which is the
probability distribution given by Eq. (1). This description of a physical
measuring procedure is not the most detailed one, and any particular
identity resolution can generally be realized by various different 
physical devices.

A special case of POVM is provided by the spectral orthogonal identity 
resolutions generated by the families of spectral projectors associated
with the self-adjoint operators in ${\cal H}$. These identity resolution 
satisfy the additional requirement
\begin{displaymath}
M_t(\Delta_1)M_t(\Delta_2)=0,\quad  \mbox{if} \quad \Delta_1\cap \Delta_2=0.
\end{displaymath}
It should be noted that this approach completely
ignores the problem of finding the state of the system after the
measurement which gave a particular result. However, since in this paper
we shall not be interested in the system state after the measurement, 
it will be sufficient for us to stay at this simplest level of the
measurement procedure so that we do shall not use the concept 
of instrument (or superoperator) [4,11--15].

The states can be reliably distinguished with the measurement described by the 
following orthogonal identity resolution in ${\cal H}$
\begin{equation}
{\cal P}_0+{\cal P}_1+{\cal P}_{\bot}=I,\quad {\cal P}_{0,1}=
|\psi_{0,1}\rangle\langle\psi_{0,1}|,\quad
{\cal P}_{\bot}=I-{\cal P}_0-{\cal P}_1,
\end{equation}
where ${\cal P}_{0,1}$ are the projectors on the subspace ${\cal H}_{0,1}$ 
spanned on the states $|\psi_{0,1}\rangle$, and ${\cal P}_{\bot}$ is the 
projector on the subspace 
${\cal H}_{0,1}^\bot=({\cal H}_0\oplus{\cal H}_1)^{\bot}$. 
For example, for the input state $|\psi_0\rangle$ the probability of obtaining 
the outcome in the channel $0$ (the outcome space is $\Theta=\{0,1,\bot\}$) is
\begin{equation}
\mbox{Pr}\{|\psi_0\rangle\}=\mbox{Tr}\{|\psi_0\rangle\langle\psi_0|
{\cal P}_0\}=1,
\end{equation}
while the probability for obtaining an outcome in the channels 
${\cal P}_{1,\bot}$ is identically zero:
\begin{equation}
\mbox{Pr}\{|\psi_0\rangle\}=\mbox{Tr}\{|\psi_0\rangle\langle\psi_0|
{\cal P}_{1,\bot}\}=0.
\end{equation}
Similar relations hold for the input state $|\psi_1\rangle$. 
Equations (2--4) mean that the orthogonal states are reliably distinguishable.
An important point is that the duration of the considered measurement
procedure has not yet been mentioned. 

{\it Our conclusions concerning the
time moment when the final result is obtained will be related to the
time required for the result to be ``communicated'' to the observer; to be more
precise, the estimates that will be obtained below provide the lower bound on 
this time interval. To derive these estimates, one should only know the
identity resolution $M_t(\Delta)$ rather than the instrument (superoperator) 
corresponding to the particular measurement procedure. The outcome space
$\Delta$ can have arbitrarily complicated nature, but in any case the
complete description of the measurement procedure involves description
of the spatial domains involved (either explicit or implicit) which proves
to be sufficient for the derivation of the restrictions imposed by the
finiteness of the speed of light.}

In the non-relativistic case, specification of the instrument (superoperator)
allows one not only to calculate the probabilities of different measurement
outcomes but also to determine the state of the quantum system after the
measurement which gave a particular outcome. The relativistic quantum theory
still lacks a clear and consistent description of the state of quantum
system just after the measurement procedure (i.e. the system state vector 
collapse). Various aspects of this problem were discussed in the papers 
by Hellwig and Krauss [16], Aharonov and Albert [17], Ghirardi {\it et al.} [18], 
and Finkelstein [19]. A comprehensive description of the measurement 
procedure should be given by a complete relativistic theory of quantum  
measurements which is yet to be biult. However, we believe that some 
problems can be solved even without a detailed description of the
measurement procedure.

Up to this moment, we have only used the properties of the abstract
Hilbert state space of the quantum system. Therefore, taking into account
that the states of the relativistic quantum fields are described by 
the rays in the Hilbert space state, one can conclude that the measurements
over quantum fields are also described by identity resolutions.
Formally, the specific realization of the abstract Hilbert space does 
not matter and in each case can be chosen in the form most suitable for
a particular problem considered. However, all quantum states should 
be associated with some physical system and all the measurements and other 
manipulations with the quantum systems are performed in space (or space-time
in the relativistic case). There exist no physical systems with the degrees 
of freedom decribed by the state vector in a Hilbert state space separately
from the spatial degrees of freedom (because actually the different sorts
of particles are classified according to the irreducible representations of 
the Poincare group containing the subgroup of translations in the Minkowskii
space-time [7]).

In the non-relativistic quantum mechanics, due to 
the absence of any restrictions on the maximum possible speed, the observer
can in principle to instantaneously (at an arbitrarily chosen moment of time)
obtain the results of non-local measurements even for infinitely extended
states. In the relativistic quantum field theory the situation is quite different.
First of all, the quantum field states are generated by the field operators
(more accurately, by the operator-valued distributions) [7]. Smearing functions 
in the momentum representation are defined by their values on the mass shell
resulting in the fundamental non-localizability in the position space, i.e.
the corresponding smearing function supports in the position spaces are 
unbounded sets [7,20--24]. However, one can construct the free field states
whose spatial localization is arbitrarily close to the exponential one
($\sim \mbox{exp}(-\alpha|{\bf x}|/\mbox{ln(ln(}\ldots|{\bf x}|)))$, with
any $\alpha$). In the absence of the limitations on the maximum possible
speed this fact alone would not result in any restrictions, just as it 
is the case for the non-relativistic quantum mechanics. However, in the
presence of the maximum allowed propagation speed for both quantum and 
classical objects the non-localizability (which itself arises when the
special relativity requirements are taken into account in the field 
quantization procedure [7]) results in a situation which is quite different 
from the non-relativistic case. Reliable distinguishing of a pair of 
orthogonal states of a quantum field requires the access to the entire 
space and hence the time required for the measurement outcome to be 
conveyed to the observer is infinite. However, the answer of the sort
that distinguishing between the two orthogonal states of a quantum field
requires infinite time is physically hardly satisfactory.

The formulation of the problem where the observer has the access to the entire 
space to reliably (with unit probability) distinguish between the two states
is hardly sensible. The observer can never control the entire space.
Therefore, the requirement of reliable distinguishability should be weakened
and reformulate the problem in the following way. The observer controls 
a finite (although arbitrarily large) spatial domain where he can perform
any measurements. Our aim is to find relation between the probability error 
in the state identification and the domain size (or, equivalently, with the 
time $T$ required for the final result to be obtained by the observer) and
the structure of the states themselves. In other words, for the specified
input states and domain size (and therefore the time required to obtain the 
measurement outcome) one has to find the optimal measurement minimizing
the state identification error probability.

We shall consider the most interesting for applications case of the gauge 
field, i.e. the photons. The electromagnetic field operators are written as [24]
\begin{equation}
A^{\pm}_{\mu}(\hat{x})=\frac{1}{(2\pi)^{3/2}}\int\frac{d{\bf k}}{\sqrt{2k_0}}
\mbox{e}^{\pm i \hat{k}\hat{x}} e_{\mu}^{m}({\bf k})a^{\pm}_{m}({\bf k})
\end{equation}
and satisfy the commutation relations
\begin{equation}
[A^{-}_{\mu}(\hat{x}),A^{+}_{\nu}(\hat{x}')]_{-}=ig_{\mu\nu}
D_{0}^{-}(\hat{x}-\hat{x}'),
\end{equation}
where $D_{0}^{-}(\hat{x}-\hat{x}')$ is the massless field commutator function
\begin{equation}
D^{\pm}_{0}(\hat{x})=\pm\frac{1}{i(2\pi)^3}
\int \frac{d{\bf p}}{2p_0}
\mbox{e}^{\pm i\hat{p}\hat{x}}=
\frac{1}{4\pi}\varepsilon({x_0})\delta(\hat{x}^2),\quad
\varepsilon({x_0})\delta(\hat{x}^2)\equiv
\frac{\delta(x_0-|{\bf x}|) - \delta(x_0+|{\bf x}|)}{2|{\bf x}|}.
\end{equation}
Here $A^{\pm}_{\mu}(\hat{x})$ are the creation (annihilation) operators
of the four types of photons --- two transverse, one longitudinal, and one
temporal. The longitudinal and temporal photons are actually fictitious 
and can be eliminated by introducing an indefinite metrics [24]. Our goal
is most simply achieved by using a particular gauge. We shall work in the
subspace of physical states employing the Coulomb gauge 
$A_{\mu}=({\bf A},\varphi=0)$ thus dealing with the two physical transverse
states of the electromagnetic field. The operator-valued distribution is 
a three-component vector
\begin{equation}
\vec{\mbox{\boldmath $\psi$}}(\hat{x})=\frac{1}{(2\pi)^{3/2}}
\int_{V^+_0}\frac{d{\bf k}}{\sqrt{2k_0}}
\sum_{s=\pm 1}{\bf w}({\bf k},s)\{a({\bf k},s)\mbox{e}^{-i\hat{k}\hat{x}}+
a^+({\bf k},-s)\mbox{e}^{i\hat{k}\hat{x}}\},
\end{equation}
where ${\bf w}({\bf k},s)$ is a three-dimensional vector describing the 
helicity state $s=\pm 1$,
\begin{equation}
{\bf w}({\bf k},\pm)=\frac{1}{\sqrt{2}}
[{\bf e}_1({\bf k}) \pm i{\bf e}_2({\bf k})],\quad
{\bf e}_1({\bf k})\bot {\bf e}_2({\bf k}),\quad
|{\bf w}({\bf k},s)|^2=1,
\end{equation}
and ${\bf e}_{1,2}({\bf k})$ are the orthogonal vectors normal to
${\bf k}$. The field operators satisfy Maxwell equations
\begin{equation}
\nabla\times\vec{\mbox{\boldmath $\psi$}}(\hat{x})=
-i\frac{\partial}{\partial t}\vec{\mbox{\boldmath $\psi$}}(\hat{x}),
\end{equation}
\begin{displaymath}
\nabla\cdot\vec{\mbox{\boldmath $\psi$}}(\hat{x})=0.
\end{displaymath}
The smeared field operators can be written as
\begin{equation}
\vec{\mbox{\boldmath $\psi$}}(f)=
\sum_{s=\pm 1}\int\vec{\mbox{\boldmath $\psi$}}(\hat{x},s)
f(\hat{x},s)d\hat{x}=
\end{equation}
\begin{displaymath}
\frac{1}{(2\pi)^{3/2}}
\int_{V^+_0}\frac{d{\bf k}}{\sqrt{2k_0}}
\sum_{s=\pm 1}{\bf w}({\bf k},s)\{
f({\bf k},s) \mbox{e}^{-i\hat{k}\hat{x}}a^+({\bf k},s)+
f^*({\bf k},s) \mbox{e}^{i\hat{k}\hat{x}}a({\bf k},s)\},
\end{displaymath}
where the function $f({\bf k},s)$ is defined as the restriction of
$f(\hat{k},s)$ to the mass shell
($f(\hat{k},s)$ is the four-dimensional Fourier transform of $f(\hat{x},s)$,
where $f(\hat{x},s)$ is an arbitrary function from ${\cal J}(\hat{x})$, i.e.
belongs to the space of test functions).

We shall consider the problem of distinguishing between the two single-photon
states which differ only in their helicity state. The two single-photon 
states with orthogonal helicities and the same spatial amplitude $f$ can
be written as 
\begin{equation}
\vec{|\mbox{\boldmath $\psi$}}_{0,1}\rangle=\left(
\vec{\mbox{\boldmath $\psi$}}^+(f_{0,1})\right)|0\rangle=
\frac{1}{(2\pi)^{3/2}}
\int_{V^+_0}\frac{d{\bf k}}{\sqrt{2k_0}}f({\bf k})
{\bf w}({\bf k},\pm)a^+({\bf k},\pm)\mbox{e}^{-i\hat{k}\hat{x}}|0\rangle=
\end{equation}
\begin{displaymath}
\int d{\bf x} f({\bf x},t) \vec{\psi}^+({\bf x},t,\pm)|0\rangle=
\int d{\bf x}' f({\bf x}',t') \vec{\psi}^+({\bf x}',t',\pm)|0\rangle,
\end{displaymath}
where
\begin{equation}
\vec{\psi}^+({\bf x},t,\pm)=
\frac{1}{(2\pi)^{3/2}}
\int_{V^+_0}\frac{d{\bf k}}{\sqrt{2k_0}}
{\bf w}({\bf k},\pm)a^+({\bf k},\pm)\mbox{e}^{-i\hat{k}\hat{x}},\quad
f({\bf x},t)=\int d{\bf k} f({\bf k}) \mbox{e}^{i\hat{k}\hat{x}}
\end{equation}
The state with subscript 0 contains components with different {\bf k} 
but only with ``$+$'' helicity,
while the state with subscripts 1 only the ``$-$'' helicity components. The
measurement allowing to reliably distinguish between these two states is 
described by the following orthogonal identity resolution in the one-particle 
subspace:
\begin{equation}
I=\mbox{\boldmath ${\cal P}$}_{0}+\mbox{\boldmath ${\cal P}$}_{1}+
\mbox{\boldmath ${\cal P}$}_{\bot},
\quad
\mbox{\boldmath ${\cal P}$}_{0,1}=
|\vec{\mbox{\boldmath $\psi$}}_{0,1}\rangle
\langle\vec{\mbox{\boldmath $\psi$}}_{0,1}|,
\quad
\mbox{\boldmath ${\cal P}$}_{\bot}=
I-\mbox{\boldmath ${\cal P}$}_{0}-\mbox{\boldmath ${\cal P}$}_{1},
\end{equation}
where the operator identity is
\begin{equation}
I=\sum_{s=\pm}\int \mbox{\boldmath $\cal M$}_t(d{\bf x},\pm)=
\sum_{s=\pm 1}\int_{V^+_0} {d {\bf k}}
\left({\bf w}({\bf k},s)\mbox{ }|{\bf k},s\rangle\right)
\left(\langle{\bf k},s|\mbox{ }{\bf w}({\bf k},s)\right),
\quad
|{\bf k},s\rangle=a^+({\bf k},s) |0\rangle,
\end{equation}
\begin{equation}
\mbox{\boldmath $\cal M$}_t(d{\bf x},\pm)=
\left(\int d{\bf k} 
\mbox{e}^{-i\hat{k}\hat{x} }{\bf w}({\bf k},\pm)|{\bf k},\pm\rangle
\right)
\left(\int d{\bf k}' 
\langle {\bf k}',\pm|{\bf w}({\bf k}',\pm) 
\mbox{e}^{i\hat{k'}\hat{x} }
\right)
\frac{d {\bf x}}{(2\pi)^3}
\end{equation}
The time $t$ in this identity resolution is {\it a parameter} which
has the same value for all points ${\bf x}$. As will be seen later,
this time should be interpreted as the time at which the measurement 
by a classical apparatus is performed.

The probabilities of obtaining an outcome in the channels ${\cal P}_j$ are
\begin{equation}
\mbox{Pr}_i\{|\vec{\mbox{\boldmath $\psi$}}_j\rangle\}=\mbox{Tr}
\{|\vec{\mbox{\boldmath $\psi$}}_i\rangle
\langle\vec{\mbox{\boldmath $\psi$}}_i|\mbox{\boldmath ${\cal P}$}_j\}=
|\langle \vec{\mbox{\boldmath $\psi$}}_j |
\vec{\mbox{\boldmath $\psi$}}_i\rangle|^2=
\end{equation}
\begin{displaymath}
\delta_{s,s'}\Bigl|\int\int d{\bf x}d{\bf x}'f^*({\bf x},t)
D^+_0({\bf x}-{\bf x}',t-t')f({\bf x}',t')\Bigl|^2=
\delta_{s,s'} \Bigl|\int_{V^+_0}f^{*}({\bf k})f({\bf k})
\frac{d{\bf k}}{2|{\bf k}|}\Bigl|^2= \delta_{i,j},
\end{displaymath}
The values $s,s'=+$ correspond to $i,j=0$, while $s,s'=-$ 
correspond to $i,j=1$.
There is a realtionship between the field amplitudes $f({\bf x},t)$ 
and $f({\bf x}',t')$ arising because of the causality resulting from 
the propagation effects described by the commutator function. 
This relationship actually reflects the fact that the coefficients
(amplitudes $f({\bf x},t)$) are actually the expansion coefficients 
of the state vector $\vec{|\mbox{\boldmath $\psi$}}_{0,1}\rangle$ from 
the Hilbert space in two different bases, $\vec{\mbox{\boldmath $\psi$}}^+({\bf x},t,\pm)|0\rangle$
and $\vec{\mbox{\boldmath $\psi$}}^+({\bf x}',t',\pm)|0\rangle$. 

The commutator function is the scalar product of two generalized basis vectors,
\begin{equation}
D^-_0({\bf x}-{\bf x}',t-t')=-i
\langle 0|\vec{\mbox{\boldmath $\psi$}}^-({\bf x},t,\pm)
{ }\vec{\mbox{\boldmath $\psi$}}^+({\bf x}',t',\pm)|0\rangle,
\quad t>t'.
\end{equation}

Since the field amplitude $f({\bf x},t)$ is non-localizable, 
the measurement described by Eqs. (14--16) and obtaining of a reliable
result assumes the access to the entire space (to be more precise, to the
entire space-like domain where $f({\bf x},t) \neq 0$) at time $t$.
Since the indicated space-like domain is infinite, the observer needs
an infinite time to reliably distinguish between the two orthogonal states.

Let us now consider the problem of distinguishing between the two states where 
only a finite spatial domain $\Omega$ (whose supplement to the entire space is
$\overline{\Omega}$) is accessible for the measurements. The outcome space is 
the set $\Theta=\{ (+,-)\times\Omega \bigcup ?$.
The measurement outcomes in the domain $\Omega$ are accessible to the 
observer while the outcome ? formally corresponding 
to the firing of a detecor in the domain $\overline{\Omega}$ is inaccessible. 
The measurement 
is described by an identity resolution defined on $\Theta$ and related 
to the time moment $t$:
\begin{equation}
I=I_{\overline{\Omega}}+I_{\Omega},
\quad
I_{\Omega}=
\int_{\Omega} \left( \mbox{\boldmath $\cal M$}_t(d{\bf x},+)+
\mbox{\boldmath $\cal M$}_t(d{\bf x},-) \right).
\end{equation}
Intuitively, this identity resolution corresponds to an ``evenly distributed'' 
over the domain $\Omega$  classical apparatus which at each spatial point
${\bf x}$ produces an outcome at time $t$ in one of the two 
channels (``$+$'' or ``$-$'') corresponding to two different helicities. 

Although the identity resolution (19) formally seems to be non-local 
(contains integration over the spatial domains), the outcomes themselves 
are local (the classical device fires at a particular spatial point 
producing a measurement outcome associated with that point). In the present 
case the outome space coincides with the physical position space where
all the measurements are actually preformed in contrast to the situation where
the measurement is described by the orthogonal projectors (2--4) acting
in the Hilbert state space $\cal H$ (which are also implicitly non-local in 
the physical position space through the non-locality of the state amplitudes);
in that case the outcome space consists of three outcomes (0, 1, and $\perp$)
and one cannot tell at which spatial point the detector fired.

The aim is to correctly identify the states which are randomly produced
for measurements with known {\it a priori} probabilities
$\pi_0$ and $\pi_1$ ($\pi_0+\pi_1=1$).
In the bases $\vec{\mbox{\boldmath $\psi$}}^+({\bf x},t,\pm)|0\rangle$
and $\vec{\mbox{\boldmath $\psi$}}^+({\bf x}',t',\pm)|0\rangle$ generated
by the field operators the states are written as
\begin{equation}
\vec{|\mbox{\boldmath $\psi$}}_{0,1}\rangle=
\int d{\bf x} f({\bf x},t)\vec{\mbox{\boldmath $\psi$}}^+({\bf x},t,\pm)|0\rangle=
\int d{\bf x}' f({\bf x}',t')\vec{\mbox{\boldmath $\psi$}}^+({\bf x}',t',\pm)|0\rangle,
\end{equation}

The outcomes can either occur in the accessible domain or, formally, 
in the inaccessible domain (corresponding to the outcome ``?''). 
The probability for an outcome to take place
in the inaccessible domain $\overline{\Omega}$ is
\begin{equation}
\mbox{Pr}\{\rho,\overline{\Omega}\}=
\mbox{Tr}\{\rho I_{\overline{\Omega}}\}=
\pi_0\mbox{Tr}\{\rho_0 I_{\overline{\Omega}}\}+
\pi_1\mbox{Tr}\{\rho_1 I_{\overline{\Omega}}\}=
\pi_0 p_t+\pi_1 p_t=p_t,
\end{equation}
where
\begin{equation}
p_t=
\int_{\overline{\Omega}}d{\bf x} \Bigl|p({\bf x},t)\Bigl|^2
\end{equation}
and
\begin{equation}
p({\bf x},t)=\frac{1}{(2\pi)^{3/2}}
\int \frac{d{\bf k}}{ \sqrt{2|{\bf k}|} }
f({\bf k})\mbox{e}^{i({\bf kx}-|{\bf k}|t)}
\end{equation}

This formula describes the probability for detection of photons with the 
$+$ ($\pi_0p_t$) and ``$-$''($\pi_1p_t$) helicities in the neighbourhood
of a random point $d{\bf x}$ at time $t$. The function $f({\bf x},t)$
is the state amplitude at time $t$. If the measurement is performed at a 
different time $t_1$, the contributions will be given by the points
where the state amplitude at time $t_1$ is different from zero, i.e. the
causally related points (see Eq. (18)).

Therefore, if no outcome was detected in the inaccessible domain, the observer 
should conclude that the outcome took place in the inaccessible domain
from the states $\vec{|\mbox{\boldmath $\psi$}}_0\rangle$ or
$\vec{|\mbox{\boldmath $\psi$}}_1\rangle$ with the probabilities
\begin{equation}
p_{0}=\frac{\textstyle \pi_0 p_t}{\textstyle \pi_0 p_t+\pi_1 p_t}=\pi_0
\quad {\rm and}\quad
p_{1}=\frac{\textstyle \pi_1 p_t}{\textstyle \pi_0 p_t+\pi_1 p_t}=\pi_1,
\end{equation}
respectively.

Thus, if the measurement outcome took place in the inaccessible domain,
the state identification error probability is equal to the product 
of the conditional probability of incorrect state indentification 
for the case of the ? measurement outcome  
and the fraction of the outcomes (relative frequency of these
outcomes) in the inaccessible domain $\overline{\Omega}$:
\begin{equation}
P_{e}(\overline{\Omega})=
(\textstyle \pi_0 p_{1} + \pi_1 p_{0} ) p_t .
\end{equation}
If the states are produced with equal probabilities ($\pi_0=\pi_1=1/2$), 
the error probability is imply equal to the fraction of outcomes in 
$\overline{\Omega}$. The total probability of outcomes in the entire 
outcome space $\Omega\bigcup\overline{\Omega}$ is 1 due to the 
normalization condition $\int d{\bf x}|p({\bf x},t)|^2=1$.

We shall now find the measurement minimizing the error probability for the
case where an outcome takes place in the accessible domain $\Omega$ 
(see Ref. [5] for details). One has
\begin{equation}
P_e(\Omega)=\pi_0\mbox{Tr}\{\rho_0I_{\Omega}\}+
\min_{E_0}\mbox{Tr}\{\Gamma E_0\}.
\end{equation}
In the basis consisting of the two orthogonal helicities states
($+$ and $-$) the operator $\Gamma$ has the form
\begin{equation}
\Gamma=\pi_1\rho_1-\pi_0\rho_0=
\left(
\begin{array}{cc}
\pi_1\vec{ |\mbox{\boldmath $\psi$}}_1\rangle\langle \vec{|\mbox{\boldmath $\psi$}}_1| & 0\\
0& -\pi_0\vec{|\mbox{\boldmath $\psi$}}_0\rangle\langle \vec{|\mbox{\boldmath $\psi$}}_0| \\
\end{array}
\right).
\end{equation}
The optimal measurement is easily found to be
\begin{equation}
E_0=
\left(
\begin{array}{cc}
0 & 0          \\
0 & I_{\Omega} \\
\end{array}
\right),
\quad
E_1=
\left(
\begin{array}{cc}
I_{\Omega} & 0 \\
0          & 0 \\
\end{array}
\right).
\end{equation}
According to Eqs.(26--28), the state identification error for this 
measurement is zero if the outcome takes place in the accessible domain:    
\begin{equation}
P_{e}(\Omega)=0.
\end{equation}
Therefore, making use of Eqs. (25) and (28) one obtains that the total error 
probability 
\begin{equation}
P_{e}(\Omega,\overline{\Omega})= P_{e}(\Omega)+P_{e}(\overline{\Omega})=
2\pi_0\pi_1 p_t= 2\pi_0\pi_1 \int_{\overline{\Omega}}d{\bf x}
\Bigl|p({\bf x},t)\Bigl|^2
\end{equation}
and is determined by the fraction of outcomes in the inaccessible domain.
Since the domain $\Omega$ (and hence its supplement to the entire space 
$\overline{\Omega}$) are specified in advance, the error is minimized by 
choosing the time $t$ (the moment when the measurement is performed)
corresponding to the minimal fraction of outcomes in the inaccessible domain.
This requirement is intuitively obvious since because of the spatio-temporal 
evolution of the amplitude $f({\bf x},t)$ the measurement
should be started at the moment when the squared modulus of $p({\bf x},t)$
integrated over the accessible domain reaches its maximum (accordingly, its
integral over the inaccessible domain is minimal).

To find out whether the outcome did take place in the accessible domain,
the observer should look through the entire accessible domain to check
if the measuring device (detector) fired in one of the channels (for 
``$+$'' or ``$-$'' helicities) at some point in $\Omega$ after the time $t$.
The domain $\Omega$ cannot be scanned faster than in time $T$ determined by 
the condition of covering this domain by the backward light cone (see 
preceding discussion) as required by the special relativity theory.

Thus, we have found an explicit expression for the measurement minimizing
the state identification error probability in the problem of distinguishing
between the two orthogonal states for the given spatial domain accessible
for the measurements (and, accordingly, the time $T$ required to obtain 
the final result). The time $T$ of course depends on the structure of the 
states involved. The time $T$ should in no case be interpreted as the
duration of the measurement procedure. In each particluar experiment the 
measurement result arises at a certain random point {\bf x} in the domain
$\Omega$ at time $t$. In some measurement the point {\bf x} can coincide 
with the point where the observer is located at time $t$ and in those cases
the states will be distinguished instantly ($T_{min}=0$). However, in other
experiments the outcomes will occur at other spatial points so that  
some finite time will be required for the observer to check if 
the measurement outcome did take place in the domain $\Omega$ for one of the
two helicity channels. Time interval $T$ is actually the minimum time 
allowing a relible distinguishability of the two states for the measurement
outcomes arising at arbitrary point of the domain $\Omega$.


The authors are grateful to Prof. V.L.Golo for valuable critical remarks.

This work was supported by the Russian Fund for Basic Research 
(grant N 99-02-18127),  the project  ``Physical foundations of quantum
computer'' and the program ``Advanced technologies and devices of micro- 
and nanoelectronics'' (project N 02.04.5.2.40.T.50).

{\small


\begin{thebibliography}{99}
\bibitem{1} J.von Neumann, {\it Mathematical Foundations of Quantum
   Mechanics}, Princeton Uni\-ver\-sity, Princeton, NJ, 1955.
\bibitem{2}C.H.Bennett, Phys. Rev. Lett. {\bf 68}, 3121 (1992);
C.H.Bennett, G.Brassard, N.D.Mermin, Phys. Rev. Lett. {\bf 68}, 557 (1992).
\bibitem{3}C.W.Helstrom, Information and Control, {\bf 10} 254 (1967);
{\it Quantum Detection and Estimation Theory}, Mathematics in Science and 
Engineering, vol.123, Academic Press, (1976).
\bibitem{4}A.S.Holevo, {\it Probabilistic and Statistical Aspects of
Quantum Theory}. North Holland Publishing Corporation, Amsterdam, 1982;
{\it Lectures on Statistical Structure of Quantum Theory}, (1999) pp.1--177.
\bibitem{5}C.A.Fuchs, xxx.lanl.gov/quant-ph/9601020.
\bibitem{6}L.Landau and R.Peierls, Zeits. f\"ur Phys., {\bf 69}, 56 (1931).
\bibitem{7}N.N.Bogolubov, A.A.Logunov, A.I.Oksak, I.T.Todorov,
{\it General Principles of the Quantum Field Theory},
Moscow, Nauka, 1987 (in Russian).
\bibitem{8}I.M.Gel'fand and N.Ya.Vilenkin, {\it Some Applications of
Harmonic Analysis. Rigged Hilbert Spaces (Generalized Functions,
issue 4)}, Moscow, Fizmatgiz, 1961 (in Russian).
\bibitem{9}N.Bohr and L.Rosenfeld, Math.-Fys. Medd., {\bf 12}, 3 (1933).
\bibitem{10}V.B.Berestetskiy, E.M.Lifshits, and L.P.Pitaevskiy,
{\it Quantum Electrodynamics}, Nauka, Moscow, 1982 (in Russian).
\bibitem{11}L.D.Landau and E.M.Lifshits, {\it Field theory}, Nauka, Moscow,
1973 (in Russian).
\bibitem{12} K.Kraus, {\it States, Effects and Operations},
Springer-Verlag, Berlin, 1983.
\bibitem{13}P.Busch, M.Grabowski, P.J.Lahti, {\it Operational Quantum
Physics}, Springer Lecture Notes in Physics, v.{\bf 31}, 1995.
\bibitem{14} G.L\"uders, Ann. Physik, {\bf 8} (6), 322 (1951).
\bibitem{15}M.Ozawa, J. Math. Phys., {\bf 25}, 79 (1984); {\bf 34}, 5596 (1993).
\bibitem{16}K.-E.Hellwig, K.Krauss, Phys. Rev., {\bf D1}, 566 (1970).
\bibitem{17}Y.Aharonov, D.Z.Albert, Phys. Rev., {\bf D21}, 3316 (1980);
{\bf D24}, 359 (1981); {\bf D29}, 228 (1984). 
\bibitem{18}G.C.Ghirardi, in {\it Open Systems and Measurement in Relativistic 
Theory}, eds.H.P.Breuer, F.Petruccione, Springer, (1999).
\bibitem{19}J.Finkelstein, {\it Property attribution and the projection 
postulate in relativistic quantum theory}, Phys. Lett., {\bf A},
to be published.
\bibitem{20}D.A.Kirzhnits, Usp.Fiz.Nauk, {\bf 90}, 129 (1966).
\bibitem{21}N.N.Meiman, ZhETF, {\bf 47}, 1966 (1964).
\bibitem{22}A.M.Jaffe, Phys. Rev., {\bf 158} 1454 (1967).
\bibitem{23}I.Bialynicki-Birula, Phys. Rev. Lett., {\bf 80}, 5247 (1998).
\bibitem{24}N.N.Bogolubov and D.V.Shirkov, {\it Introduction to the Quantum Field 
Theory}, Moscow, Nauka, 1973 (in Russian).
\end{thebibliography}
\end{document}